\documentclass[12pt]{article}
\usepackage{amsmath,amssymb,amsfonts}
\usepackage{graphicx}
\begin{document}

{\vspace*{2cm}
\begin{center}
{\textbf{THE MAGNETIC ORDERING IN DILUTED CRYSTAL MAGNETS WITH
INDIRECT EXCHANGE INTERACTION}}
\end{center} }
\begin{center}
 \sl V.I. Belokon$^{1)}$, K.V. Nefedev$^{2)}$, M.A. Savunov
\\
\emph{Far Eastern National University\\
690950, Russia, Vladivostok, Sukhanova str. 8-43.}
\end{center}
\begin{center}
E-mail:\hspace{0.2cm}1)
Belokon@ifit.phys.dvgu.ru\\\vspace{-0.5cm}\hspace{0.9cm} 2)
knefedev@phys.dvgu.ru
\end{center}

\begin{abstract}
 The closed system of equations for a determination of parameters
distribution function for random interaction fields was calculated
analytically.  The estimations of critical concentrations for
phase transitions in diluted crystal magnets with face centered
cubic (fcc), volume centered cubic (vcc) and simple cubic (sc)
lattices with interaction of Ruderman-Kittel-Kasua-Yosida between
spins were made.
\end{abstract}

Magnetic properties of disordered systems are investigated for a
long time. In this direction great number of works were made,
there are large number of books [1-5] and reviewed publications
[6,7]. The most of approaches to theoretical investigation of
magnetism of disordered matter use a assumption about random
distribution of exchange integrals in Hamiltonian of spin-spin
interaction, at that a parameters of distribution function must
have agreement with experimental data. We have formulated in [8,9]
several different approach, which one help to calculate a
distribution function of random fields of exchange interaction for
amorphus magnets. A parameters of such function are in agree with
each other and are calculated with using of law of exchange
interaction (or magnetic moments of particles, clusters, grains
etc.). In this article the proposed approach was extend for
investigation of magnetic properties of crystal materials.

\begin{center}
 1. DENSITY OF FIELD DISTRIBUTION
\end{center}
\vspace*{3mm}
In [8] the density of distribution of random field exchange
interaction have been calculated for amorphus ferromagnetic
material. In this case the probability of the hit of particle in
element with volume $ d V$ can be determinate as $\frac{\textstyle
 d V}{\textstyle V}$, and density of distribution of particles
over magnetic moment $\textbf{S}$
\begin{equation}
\tau(\textbf{S}) d\textbf{S}=\frac{1}{2\pi
S^{2}\sin^{2}\vartheta}\delta(S-S_{0})
[\alpha\delta(\vartheta)+\beta\delta(\vartheta-\pi)]d\textbf{S}
\end{equation}
For crystal lattices difference is in that the density of
distribution of interacting particles, which ones have same
magnetic moment $\textbf{S}_{0}$, over $\textbf{S}_k$ in the model
of Ising is
\begin{equation}
\tau_{k}(\textbf{S}_{k})d\textbf{S}_{k}=\frac{1}{2\pi
S^{2}\sin^{2}\vartheta_k}[\alpha_{k}\delta(\vartheta_{k})+\beta_{k}\delta(\vartheta_{k}-\pi)]
\left[\frac{N-N_{0}}{N}\delta(\verb"S"_{k})+\frac{N_{0}}{N}\delta(S_{k}-S_{0})\right]d
\textbf{S}
\end{equation}
Here $\alpha_{k}$ --- a relative number of particles, which ones
have orientation in ''positive'' direction, $\beta_{k}$
--- ''negative'', $\alpha_{k}+\beta_{k}=1$,
$\delta(\vartheta_{k})$
--- delta-function of Dirac, $N$ --- number of sizes of crystal lattice,
$N_{0}$ --- number of interacting particles. The characteristic
function in this case is
\begin{center}
$A(\rho)=\textstyle\displaystyle\int W(H)\exp\left(i\rho H\right)d
H=$
\end{center}
\vspace{-1cm}
\begin{equation}
=\textstyle\displaystyle\prod_{k}\left[\left(1-p\right)+
p\left(\alpha\exp\left\{i\rho\varphi_{k}(S_{0},r_{k,0})\right\}+
\beta
\exp\left\{-i\rho\varphi_{k}(S_{0},r_{k,0})\right\}\right)\right],
\end{equation}
where $p=\textstyle\displaystyle\frac{\textstyle N_{0}}{\textstyle
N}$, $\varphi_{k}(S_{0},r_{k,0})$ --- field, which one was created
in origin of coordinates by particle in cite $r_{k,0}$. If to
leave only three terms of expansion of exponent then
\begin{center}
$\ln A(\rho)\approx i(\alpha-\beta)p\rho
\textstyle\displaystyle\sum_{k}\varphi_{k}-\frac{1}{2!}p\left[1+(\alpha-\beta)^{2}p\right]\rho^2
\textstyle\displaystyle\sum_{k}\varphi_{k}\approx $
\end{center}
\vspace{-1cm}
\begin{equation}
\approx i(\alpha-\beta)p\rho
\sum_{k}\varphi_{k}-\frac{1}{2!}p\rho^2\sum_{k}\varphi_{k}.
\end{equation}
The distribution function for random field exchange interaction
$W(H)$ (it is same as in case of amorphus magnet) is ''extended''
$\delta$-function with view
\begin{equation}
W(H)=\frac{1}{\sqrt{\pi}B}\exp\left(\frac{\left[H-H_{0}(\alpha-\beta)\right]^2}{B^2}\right),
\end{equation}
\begin{center}
$H_{0}=p\textstyle\displaystyle\sum_{k}\varphi_{k}$,\hspace{1cm}
$B^2\approx2p\textstyle\displaystyle\sum_{k}\varphi_{k}^2$.
\end{center}
Remind that for amorphus magnets
\begin{center}
$H_{0}=n\textstyle\displaystyle\int_{V}\varphi dV$,\hspace{1cm}
$B_0^2\approx2n\textstyle\displaystyle\int_{V}\varphi^2 dV$,
\end{center}
where $n$ --- volume concentration.

After thermodynamical averaging of characteristics of field
''origin'' in cite
\begin{equation}
\overline{\alpha}=\textstyle\displaystyle\frac{\exp\left\{\textstyle\displaystyle\frac{S_0
H}{kT}\right\}}{2\cosh\left\{\textstyle\displaystyle\frac{S_0
H}{kT}\right\}}, \hspace{1cm}
\overline{\beta}=\textstyle\displaystyle\frac{\exp\left\{-\textstyle\displaystyle\frac{S_0
H}{kT}\right\}}{2\cosh\left\{\textstyle\displaystyle\frac{S_0
H}{kT}\right\}},
\end{equation}
\begin{equation}
\left|\overline{S}\right|=S_0\left|\tanh{\left\{\textstyle\displaystyle\frac{S_0
H}{kT}\right\}}\right|,
\end{equation}
and after configuration averaging for relative magnetization $M$
per one cite, for $<H_0>$ and $<B^2>$ easy to receive a system of
selfconsistented equations
\begin{equation}
\left\{\begin{array}{l}<M>=\displaystyle \int\tanh\left\{\displaystyle \frac{S_0 H}{k T}\right\}\;W(H)\verb"d"H\\
<H_{0}>=p\textstyle\displaystyle\sum_k\varphi_k\displaystyle\int\left|\tanh\left\{\textstyle\displaystyle\frac{S_0
H}{kT}\right\}\right|W(H)\verb"d"H
\\<B^2>=2p\textstyle\displaystyle\sum_k\varphi_k^2\displaystyle\int\tanh^2\left\{\textstyle\displaystyle\frac{S_0
H}{kT}\right\}W(H)\verb"d"H.
\end{array}\right.
\end{equation}
This system is analog of Sherrington-Kirkpatrick's system of
equations, see for example [10,11]. Here $B$ --- parameter of
order (similar to $q$), and in case of $M=0$ it describes a spin
glass state. The advantage of system (8) is in that the basic
parameters of distribution function ($<H_0>$, $<B^2>$) are
connected with each other and depend of law of interaction of
particles. Here and future for $<H_0>$ and $<B^2>$ we shall have
left out a sing of averaging.

The system (8) can be essentially simplified, if to make a
substitute of distribution function
\begin{equation}
W(x)=\textstyle\displaystyle\frac{1}{\sqrt{\pi}B}\textstyle\displaystyle\exp
\left\{-\textstyle\displaystyle\frac{ x^2}{B^2}\right\}
\end{equation}
by rectangular
\begin{equation}
\widetilde{W}(x)=\left\{\begin{array}{l}0,\hspace{1cm} -{B}> x,\hspace{0.5cm} {B}<x\\
\displaystyle\frac{1}{\displaystyle 2{B}}\hspace{7mm}
-{B}\leqslant x\leqslant{B}.\end{array}\right.
\end{equation}

There are example of numerical solving of equation for $M$ with
exact and approximated functions [9]. Near of points of phase
transition (small $M$ and $B$) error in calculations is
negligible. The available estimations of critical density $p_c$
are addition argument in favour of this substitution. $p_c$
corresponds to percolation threshold. It can be received if to
consider of system (8) for case of direct exchange.

So, from equation
\begin{equation}
{M}=\frac{1}{2B}\textstyle\displaystyle\int_{-B}^{B}\tanh\left\{\displaystyle
\frac{S_0 (H+H_0)}{k T}\right\}\;\verb"d"H
\end{equation}
it follows that a Curie point can be determined from relation
\begin{equation}
\frac{H_0}{B}\tanh\left\{\displaystyle\frac{S_0 B}{k T}\right\}=1.
\end{equation}
It is obviously that condition
$\textstyle\displaystyle\frac{H_0}{B}=1$ defines a maximal
concentration $p_c$, below of which a ordering does not exist even
at $T=0$. For direct exchange $\varphi_k=f_0$, and summation must
goes over nearest neighbors.
\begin{center}
$\displaystyle\gamma=\frac{H_0}{B_0}=\displaystyle\frac{\displaystyle
p_c\displaystyle\textstyle z f_0}{\sqrt{2p_c z}f_0}=1$,
\end{center}
\vspace{-1cm}
\begin{equation}
p_{c}={\frac{\textstyle 2}{\textstyle z}},
\end{equation}
where $z$ --- number of nearest neighbors.

The critical concentrations $p_c$, calculated by help of formula
(13), have good agreement with knowing results in theory of
percolation [12-15].

{
\begin{center}
2. RKKY INTERACTION
\end{center} }
\vspace*{3mm}
For RKKY interaction the ''strength of exchange field'' is
\begin{equation}
\varphi=A F(x),
\end{equation}
where $x=2k_F R$, $k_F$ --- impulse of free electron on Fermi
surface,
\begin{equation}
F(x)=\frac{x \cos{x}-\sin x}{x^4},
\end{equation}
and $A$ has dimension of magnetic field and defines of intensity
of exchange interaction.

Since for ''standard'' metal $k_F^3=3\pi^2n_s$, where $n_s$
--- concentration of free electrons, Fermi-impulse is comparable
with parameter of lattice ($k_F\sim a$). The results of summation
in formulas for $H_0$ and $B^2$ can have significant dependence
from mutual disposition of interacting atoms, i.e. from type of
crystal lattice. It is known that distance between atoms for
different lattices can be calculated as

1) SC
\begin{equation}
R_{n_1,n_2,n_3}=a\sqrt{n_1^2+n_2^2+n_3^2},
\end{equation}

2) FCC
\begin{equation}
R_{n_1,n_2,n_3}=\frac{a}{2}\sqrt{(n_1+n_2)^2+(n_1+n_3)^2+(n_2+n_3)^2},
\end{equation}

2) VCC
\begin{equation}
R_{n_1,n_2,n_3}=\frac{a}{2}\sqrt{(n_1+n_2-n_3)^2+(n_1+n_3-n_2)^2+(n_2+n_3-n_1)^2},
\end{equation}
where $n_1$, $n_2$ and $n_3$ --- whole numbers. The number of
atoms, which ones have same distances, equals to number of integer
roots of equations (16-18).

The parameters of distribution function $H_0$ and $B^2$ for
crystal magnetic alloys, which ones have given type of lattice,
can be calculated by summation:
\begin{equation}
H_0=pA\sum_{n_1,n_2,n_3} F(2k_F R_{n_1,n_2,n_3}),
\end{equation}
\begin{equation}
B^2=2pA^2\sum_{n_1,n_2,n_3} F^2(2k_F R_{n_1,n_2,n_3}).
\end{equation}
At the calculation of $H_0$ and $B^2$ it is necessary to have in
view, that concentration of free electrons
$n_s=\textstyle\displaystyle\frac{4}{a^3}$ for FCC lattice, for
VCC $n_s=\textstyle\displaystyle\frac{2}{a^3}$ and for SC lattice
$n_s=\textstyle\displaystyle\frac{1}{a^3}$.

The numerical estimations of parameters of distribution function
$W(H)$, which ones have been made with taking in account a first
20 roots of equations (16-18), when sums (19-20) go to suturation,
for FCC, VCC and SC lattice, correspondingly, are:

FCC: $H_0=0.037pA$, \hspace{1cm}
$B=0.012\sqrt{p}A$,\hspace{1cm}$\textstyle\displaystyle\frac{H_0}{B}=3.14\sqrt{p}$\hspace{0.1cm};

VCC: $H_0=0.032pA$, \hspace{1cm}
$B=0.011\sqrt{p}A$,\hspace{1cm}$\textstyle\displaystyle\frac{H_0}{B}=2.83\sqrt{p}$\hspace{0.1cm};

SC: \hspace{0.3cm}$H_0=0.011pA$, \hspace{1cm}
$B=0.016\sqrt{p}A$,\hspace{1cm}$\textstyle\displaystyle\frac{H_0}{B}=0.69\sqrt{p}$\hspace{0.1cm}.

From this data it follows that for FCC lattice in case of RKKY
exchange in standard metal the ferromagnetic state is possible at
$p>p_c\cong0.32$, and the spin glass ordering takes place at
$p<p_c$. For VCC the ferromagnetism is possible only at
$p>p_c\cong0.35$. And for SC it is possible only the spin glass
state at low temperature and the paramagnetism at high one.

Thus, the method of random field exchange interaction allows to
make a estimations of critical concentrations of phase transitions
in diluted magnets in dependence of crystal structure and of type
of exchange interaction.

This work was made at financial support of Ministry of Education
and Science of Russian Federation, and at mutual grant of Ministry
of Education and Science of Russian Federation and DAAD, program
M. Lomonosov (ref: 325-A/05/05416).

\end{document}